\documentclass[aps,twocolumn,preprintnumbers,prb,amsmath,
amssymb,amsfonts,superscriptaddress,floatfix]{revtex4}

\usepackage{amsfonts}
\usepackage{amssymb}
\usepackage{amsmath}
\usepackage{ulem}
\usepackage{color}
\usepackage{latexsym}
\usepackage{graphicx}
\usepackage{graphics}
\usepackage{floatflt}
\usepackage{epsfig}
\usepackage{overpic}
\usepackage{soul} 
\usepackage{varioref,xr-hyper}
\usepackage[latin1]{inputenc}
\newcommand{\be}{\begin{equation}}
\newcommand{\ee}{\end{equation}}
\newcommand{\bea}{\begin{eqnarray}}
\newcommand{\eea}{\end{eqnarray}}

\def\nn{\nonumber}
\def\lb{\label}
\def\pref#1{(\ref{#1})}

\def\ra{\rightarrow}

\def\Ra{\Rightarrow}

\def\bk{{\bf k}}

\def\bq{{\bf q}}

\def\bQ{{\bf Q}}

\def\bS{{\bf S}}

\def\e{\varepsilon}

\def\l{\lambda}

\def\s{\sigma}

\def\G{\Gamma}
\def\D{\Delta}

\begin{document}
\title{Nematic Pairing from Orbital Selective Spin Fluctuations in FeSe}
\author{Lara Benfatto}
\affiliation{ISC-CNR and Department of Physics, ``Sapienza'' University of Rome, P.le
A. Moro 5, 00185, Rome, Italy} 
\author{Bel\'en Valenzuela}
\affiliation{Materials Science Factory, Instituto de Ciencia de Materiales de Madrid, 
ICMM-CSIC, Cantoblanco, E-28049 Madrid, Spain}
\author{Laura Fanfarillo}
\affiliation{CNR-IOM and International School for Advanced Studies (SISSA), Via
Bonomea 265, I-34136, Trieste, Italy}
\email{laura.fanfarillo@sissa.it}
\date{\today}


\begin{abstract} {FeSe is an intriguing iron-based superconductor. It presents an unusual nematic state
without magnetism and can be tuned to increase the critical superconducting
temperature. Recently it has been observed a noteworthy anisotropy of the
superconducting gaps. Its explanation is intimately related to the understanding
of the nematic transition itself. Here we show that the spin-nematic scenario
driven by orbital-selective spin-fluctuations provides a simple scheme to
understand both phenomena. The pairing mediated by anisotropic spin
modes is not only orbital selective but also nematic, leading to stronger pair
scattering across the hole and $X$ electron pocket. The delicate balance between
orbital ordering and nematic pairing points also to a marked $k_z$ dependence of
the hole-gap anisotropy.}\end{abstract}

{\bf \maketitle }

\section*{INTRODUCTION}
Soon after the discovery of superconductivity in iron-based systems it has been 
proposed that pairing could be unconventional, i.e. based on a non-phononic 
mechanism\cite{Mazin_PRL2008,Kuroki_PRL2008}. This proposal has been triggered, 
from one side, by the small estimated value of the electron-phonon coupling, 
and, from the other side, by the proximity in the temperature-doping phase 
diagram of a magnetic instability nearby the superconducting (SC) one. Within an 
itinerant-electron picture pairing could be provided by repulsive 
spin-fluctuations (SF) between hole and electron pockets, connected by the same 
wavevector characteristic of the spin modulations in the magnetic phase (see 
Fig.\ 1). This suggestion has been supported and confirmed by an 
extensive theoretical work, aimed from one side to establish why inter-pockets 
repulsion can overcome the intra-pocket one\cite{Chubukov_Review2012} and from 
the other side to provide a quantitative estimate of the SC properties starting 
from RPA-based description of the SF susceptibility\cite{Platt_AdvPhys2013, 
Hirschfeld_Review2016}.
 
The success of the itinerant scenario as a unified description of Fe-based 
materials has been partly questioned by the discovery of superconductivity in 
the FeSe system. Recent experiments\cite{Rahn_PRB2015, Wang_NatMat2016, 
Wang_NatCom2016, He_PRB2018, Wiecki_PRB2018} detected sizeble SF in FeSe, 
however, a magnetic phase appears only upon doping. 
Superconductivity emerges below $T_c \sim 9$ K from the so-called nematic 
phase\cite{Coldea_Review2018}. Here at temperatures below $T_S=90$ K the 
anisotropy of the electronic properties is far larger than what expected across 
a standard tetragonal-to-orthorhombic transition, suggesting that it is driven 
by electronic degrees of freedom\cite{Gallais_Review2016, Coldea_Review2018}. In 
particular, ARPES experiments clearly show a dramatic change of the Fermi 
surface (FS) across $T_S$, that can be reproduced with an effective 
crystal-field splitting of the various orbitals\cite{Shimojima_PRB2014, 
Nakayama_PRL2014, Watson_PRB2015, Zhang_PRB2015, Suzuki_PRB2015, Zhang_PRB2016, 
Fanfarillo_PRB2016, Fedorov_SciRep2016, Watson_NJP2017}. 
\begin{figure}[b]
\includegraphics[angle=0,width=0.94\linewidth]{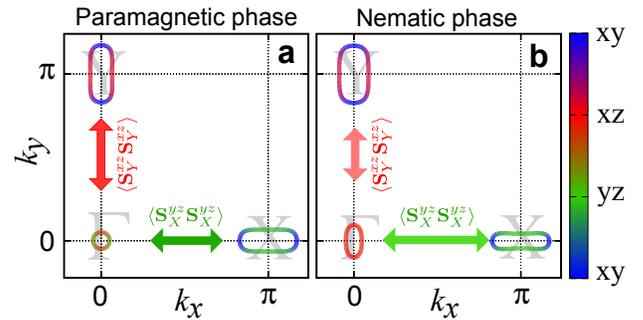} \caption{FeSe Fermi 
surfaces at $k_z = 0$. ({\bf a}) Paramagnetic phase. ({\bf b}) Nematic phase. 
The colors represent the main orbital character of the Fermi surface. The 
green/red arrows denote the orbital selective spin fluctuations (OSSF), 
connecting hole and electron pockets at different momenta, see Eq. 
\pref{corrx}-\pref{corry}. The spin-fluctuations along $\G X$ and $\G Y$ are 
equivalent in the paramagnetic phase  ({\bf a}) and become anisotropic in the 
nematic one  ({\bf b})} 
\label{fig1}
\end{figure}

In this situation, the explanation of the observed anisotropy of the SC gaps in 
FeSe becomes intimately related to the understanding of the nematic transition 
itself. Extensive experimental studies on FeSe-based material, ranging from 
quasiparticle interference imaging\cite{Sprau_Science2017, Song_Science2011} and ARPES 
measurements\cite{Xu_PRL2016, Hashimoto_NatCom2018, Rhodes_arxiv2018, Liu_PRX2018, 
Kushirenko_PRB2018}, to thermal probes\cite{Sato_PNAS2018, Sun_PRB2017}, suggest 
that the SC gap in FeSe is highly anisotropic on both hole and electron pockets. 
By defining $\theta$ the angle formed with the $k_x$ axis measured with respect 
to the center of each pocket, one finds that the gap is larger at $\theta=0$ on 
the $\Gamma$ pocket, where the predominant character in the nematic phase is 
$xz$\cite{Fanfarillo_PRB2016, Rhodes_arxiv2018, Liu_PRX2018}, and at $\theta=\pi/2$ on 
the $X$ pocket, where the dominant character is $yz$, Fig.\ 1b. 
Thus, accounting for an orbital-dependent SC order parameter does not 
reproduce the observed gap hierarchy, and additional phenomenological 
modifications of the pairing mechanism must be introduced\cite{Kreisel_PRB2017, 
Sprau_Science2017, Rhodes_arxiv2018} to describe the experiments.

Among the various attempts to theoretically understand the nematic phase from 
microscopic models, we have recently emphasized  the outcomes of a theoretical 
approach which correctly incorporates the feedback between orbital degrees of 
freedom and SF\cite{Fanfarillo_PRB2016,Fanfarillo_PRB2018,Fernandez_arxiv2018}. 
From one side the degree of orbital nesting between hole and electron pockets is 
crucial to determine the temperature scale where SF beyond RPA drive the 
spin-nematic instability\cite{Fanfarillo_PRB2018}, making SF at 
$\bQ_X=(\pi/a,0)$ and $\bQ_Y=(0,\pi/a)$ anisotropic below 
$T_S$\cite{Fernandes_NatPhys2014}. From the other side SF renormalize the 
quasiparticle dispersion, so that the orbital ordering observed below $T_S$ is a 
consequence of the spin nematicity, thanks to an orbital-selective shrinking 
mechanism\cite{Fanfarillo_PRB2016}. In this work we show that such 
orbital-selective spin fluctuations (OSSF) provide also the key pairing 
mechanism needed to understand the SC properties of FeSe. Within an 
orbital-selective spin-nematic scenario, the $C_4$ symmetry breaking of the SF 
below $T_S$ provides a pairing mechanism that is not only orbital selective but 
also nematic, in the sense that inter-pocket pair scattering along the $\Gamma 
X$ and $\Gamma Y$ directions becomes anisotropic. As we show below, accounting 
only for the nematic band-structure reconstruction of the FS, the SC gap of the 
$\Gamma$ pocket follows the modulation of the dominant $xz$ orbital, with a weak 
relative maximum at $\theta=\pi/2$, in striking disagreement with the 
experiments. The nematic pairing provided by OSSF is crucial to enhance the $yz$ 
component of the SC order parameter, explaining why the anisotropy of the SC gap 
at $\Gamma$ follows the subdominant $yz$ orbital character of the underlying 
Fermi surface\cite{Rhodes_arxiv2018, Liu_PRX2018}. We also discuss its 
implications for the gap-structure measured at $k_z=\pi$ (ref.s 
\onlinecite{Xu_PRL2016, Hashimoto_NatCom2018, Kushirenko_PRB2018}), where hole 
pocket retains a larger $yz$ character even in the nematic phase, making the 
nematic pairing responsible for an enhancement of the moderate gap anisotropy 
triggered already by  orbital-ordering effects\cite{Kang_PRL2018}.

\section*{RESULTS}

\subsection*{Model}

To compute the SC properties of FeSe we start from a low-energy 
model adapted from\cite{Cvetkovic_PRB2013}. The orbital content of each pocket is 
encoded via a rotation from the fermionic operators $c_{xz}, c_{yz}, c_{xy}$ in 
the orbital basis to the ones describing the outer hole pocket ($h$) at $\Gamma$ 
and the electronic pockets  at $X$ ($e_X$) and at $Y$ ($e_Y$): 
\bea
\lb{hp}
h_\bk&=&u_{\Gamma,\bk} c_{yz,\bk}-v_{\Gamma,\bk}c_{xz,\bk},\\
\lb{ex}
e_{X,\bk} &=& u_{X,\bk} c_{yz,\bk} - iv_{X,\bk} c_{xy,\bk}, \\
\lb{ey}
e_{Y,\bk} &= &u_{Y,\bk} c_{xz,\bk} - iv_{Y,\bk} c_{xy,\bk},
\eea
where the explicit definition of the coefficients $u_{\ell,\bk}, v_{\ell,\bk}$ 
with $\ell=\G, X, Y$ is given in Supplementary Note 1. For example, for the hole 
pocket in the tetragonal phase $u_{\G,\bk_F}\sim \cos\theta$ and 
$v_{\G,\bk_F}\sim \sin \theta$, accounting for the predominant orbital character 
of the FS represented in Fig.\ 1a. By 
using the identities \pref{hp}-\pref{ey} one can establish\cite{Fanfarillo_PRB2015, 
Fanfarillo_PRB2018} (see also Supplementary Note 2)
a precise correspondence between the 
orbital character of the spin operator and the momenta $\bQ_X$ or $\bQ_Y$ 
connecting the hole and the $X/Y$ pockets: 
\bea
\lb{sx}
 \bS \, (\bQ_X)  \equiv \bS_{X}^{yz}&=& \sum_\bk u_{\G,\bk}h_{\bk}^{\dagger} 
\, \vec{\s} \, u_{X,\bk+\bQ_X} e_{X,\bk+\bQ_X}, \\
\lb{sy}
\bS \, (\bQ_Y) \equiv  \bS_{Y}^{xz}  &=& \sum_\bk -v_{\G,\bk}h_{\bk}^{\dagger} 
\, \vec{\s} \, u_{Y,\bk+\bQ_Y} e_{Y,\bk+\bQ_Y}.
\eea
Since $xz$ states are absent at $X$ the $S^{xz}_\bq$ operator has 
no component at the wavevector $\bQ_X$ connecting the $\G$ and $X$ pocket, and 
viceversa for the $yz$ states. This leads to OSSF at different momenta, as depicted 
in Fig.\ 1: 
\bea
\lb{corrx}
\langle \bS \cdot \bS \rangle (\bQ_X) &\Ra& \langle \bS^{yz}_{X}\cdot \bS^{yz}_{X}\rangle,\\
\lb{corry}
\langle \bS\cdot \bS\rangle (\bQ_Y) &\Ra& \langle \bS^{xz}_{Y}\cdot \bS_{Y}^{xz}\rangle.  
\eea

The existence of  OSSF provides a natural explanation of the orbital ordering 
observed in the nematic phase of FeSe. In fact, the self-energy corrections due 
to spin exchange imply a shift in the chemical potential with opposite sign for 
the hole and electron pockets, leading in both cases to a shrinking of the 
FS\cite{Ortenzi_PRL2009, Fanfarillo_PRB2016} that explains why experimentally 
they are always smaller than LDA 
predictions\cite{Coldea_PRL2008,Brouet_PRL2013,Fanfarillo_PRB2016}. Within the 
OSSF model, due to the orbital-selective nature of SF, this mechanism is also 
orbital dependent\cite{Fanfarillo_PRB2016}. As a consequence, within a 
spin-nematic scenario, the $C_4$ symmetry breaking of SF along $\G X$ and $\G Y$ 
explains also the orbital ordering observed in the nematic phase. 
It has been shown\cite{Fanfarillo_PRB2016} that, by assuming stronger SF at 
$\bQ_X$ below $T_S$, the self-energy difference $\Delta\Sigma$ between  $xz$ and 
$yz$ and orbitals induced an orbital splitting being positive at $\G$ and 
negative at the electron pockets, leading to the observed deformations of the FS 
below $T_S$\cite{Suzuki_PRB2015, Watson_PRB2015, Fanfarillo_PRB2016, 
Fedorov_SciRep2016, Watson_NJP2017, Coldea_Review2018}. Even though this 
orbital-selective shrinking mechanism is generic, its effect can be 
quantitatively different in the various family of iron-based superconductors. 
For example, in the 122 family the survival of the inner hole pocket enhances 
the degree of orbital nesting between hole and electron pockets favoring 
magnetism, this explains why in 122 the nematic transition is immediately 
followed by the magnetic one\cite{Fanfarillo_PRB2018}. The quantitative 
determination of the nematic splitting induced by the nematic spin modes 
requires a direct comparison with the low-energy band dispersion, as done 
explicitly for FeSe in\cite{Fanfarillo_PRB2016}. Here we take these results for 
granted and we start from a low-energy model that includes already the effective 
masses, isotropic shrinking and nematic splittings needed to reproduce the 
ARPES FS measured in the nematic phase above $T_c$, and the $k_z$ dependence of 
the hole pocket between the $\G$ ($k_z=0$) and Z ($k_z=\pi$) point (see 
Supplementary Note 3). 
The resulting FS at $k_z=0$ is shown in Fig.\ 1. 

\begin{figure}[t]
\includegraphics[angle=0,width=0.98\linewidth]{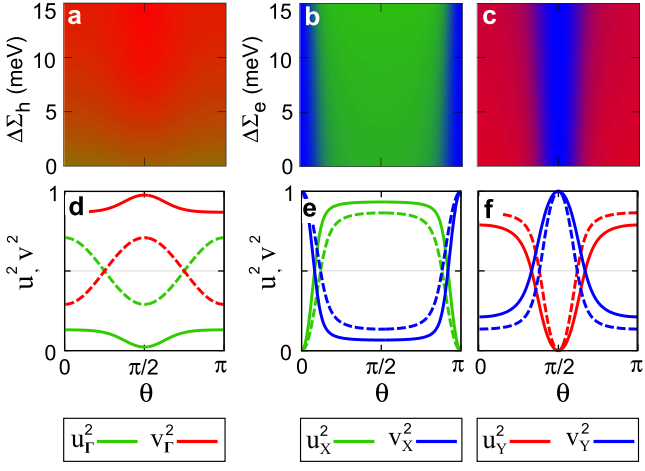}
\caption{Orbital content of the FS. ({\bf a}-{\bf c}) Color maps of the 
orbital content of the $\Gamma$\ ({\bf a}), X\ ({\bf b}), Y\ ({\bf c)} pockets FS 
as a function of the angle and of the nematic splitting $\Delta \Sigma_{h/e}$. 
The color code is the same as in Fig.\ 1. ({\bf d}-{\bf f}) 
Orbital content of the same pockets as a function of $\theta$ at $\Delta 
\Sigma_{h/e}=0$, i.e. in the tetragonal phase (dashed lines) and in the nematic 
phase $\Delta \Sigma_{h/e}=15$ meV (solid lines).}
\vspace{-0.4cm}
\label{fig2}
\end{figure}

The effect of the nematic orbital splitting on the orbital factors 
below $T_S$ is shown in Fig.\ 2. Here, 
$\Delta\Sigma_h=(\Sigma^\G_{xz}-\Sigma^\G_{yz})/2$ denotes the nematic splitting 
at $\G$ and $\Delta\Sigma_e=(\Sigma^X_{yz}-\Sigma^Y_{xz})/2$ is the nematic 
splitting at $M=(X,Y)$ with $\Sigma^\ell_{yz/xz}$ being the  $yz/xz$ orbital 
component of the real part of the self-energy for the $\ell$ pocket 
(see Supplementary Note 1). 
The maximum values are chosen to match the experimental 
ones\cite{Fanfarillo_PRB2016, Fedorov_SciRep2016, Watson_NJP2017, 
Coldea_Review2018}, i.e. $\Delta\Sigma_{h/e}\simeq 15$ meV. The most dramatic 
changes due to the nematic order are found in the orbital occupation of the hole 
pocket Fig.\ 2a,d. The presence of a relatively large spin-orbit 
coupling ($\simeq 20$ meV) implies a mixing of the  $xz$ and $yz$  orbitals on 
all the FS. However, below $T_S$ the $yz$ character of the hole pocket is 
strongly suppressed, and the pocket acquires a dominant $xz$ character even at 
$\theta=0$, as observed by the polarization dependent ARPES 
measurements\cite{Suzuki_PRB2015, Fanfarillo_PRB2016, Rhodes_arxiv2018, 
Liu_PRX2018}. At the same time the nematic splitting enhances the $yz$ 
occupation at $X$ (Fig.\ 2b,e), and suppresses the $xz$ at $Y$ (Fig.\ 
2c,f). As a consequence, one easily understands that considering the 
orbital character of the SC order parameter is not enough to explain the 
observed gap hierarchy.  In fact, on the $X$ pocket the gap is maximum at 
$\theta=\pi/2$, where the band has strong $yz$ character, while on the $\G$ 
pocket it is larger at $\theta=0$, where a dominant $xz$ character is found. The 
crucial ingredient required to account for the SC properties of FeSe comes 
indeed from the nematic pairing provided by OSSF, as we show below. 

By building up the spin-singlet vertex mediated by the SF 
\pref{corrx}-\pref{corry} one obtains (see Supplementary Note 2) a 
pairing Hamiltonian involving only the $xz/yz$ orbital sector: 
\bea
H^{xz,yz}_{\text{pair}} &=& - g_X \sum_{\bk, \bk^{\prime}} u^2_{\G, \bk} 
h^\dagger_\bk h^\dagger_{-\bk}  u^2_{X, \bk^\prime}  e_{X,-\bk^\prime} e_{X,\bk^\prime}  \nn \\
&-& g_Y \sum_{\bk, \bk^{\prime}} v^2_{\G, \bk} h^\dagger_\bk h^\dagger_{-\bk}  
u^2_{Y, \bk^\prime}  e_{Y,-\bk^\prime} e_{Y,\bk^\prime}+ h.c.
\lb{Hsc} 
\eea
The coefficients $u_{\ell,\bk}$, $v_{\ell,\bk}$, accounting for the pockets 
orbital character, preserve the $C_4$ band-structure symmetry above $T_S$ and 
reproduce the nematic reconstruction below $T_S$. The $g_{X/Y}$ couplings 
control the strengh of the pair hopping between the $\G$ and $X/Y$ pockets. 
Within a spin-nematic scenario, OSSF  below $T_S$ are stronger along $\Gamma X$ 
than along $\G Y$ leading to a {\it nematic pairing anisotropy} with $g_X>g_Y$. 
Within the present itinerant-fermions picture the SF are peaked at the 
wavevectors connecting hole-like with electron-like pockets. Thus, due to the 
absence in FeSe of the hole-like $xy$ band at $\G$ one can neglect the 
spin-mediated pairing in the $xy$ channel. However, SF at RPA level were 
found\cite{Kreisel_PRB2017} to be most prominent at $\bQ=(\pi,\pi)$. While this 
could be consistent with inelastic neutron scattering measurements at high 
temperatures, it does not account for the predominance of stripe-like SF at 
$(\pi,0)$ in the nematic phase\cite{Wang_NatMat2016}. In addition, a predominant 
$\bQ=(\pi,\pi)$ pairing channel implies a maximum gap value on the $xy$ sector 
of the electron pocket, that is in sharp contrast with the experiments. This led 
the authors of ref.s \cite{Sprau_Science2017, Kreisel_PRB2017} to 
phenomenologically introduce orbital-dependent spectral weights to suppress this 
channel (see Discussion section). In general, one can still expect that a 
smaller pair hopping between the $X, Y$ pockets is present in the $xy$ sector. 
For the sake of completeness, and with the aim of reducing the number of free 
parameters, we considered also in this case only an interband $xy$ pairing term, 
acting between the two electron-like pockets: 
\be
\lb{Hxy} 
H^{xy}_{\text{pair}} = -g_{xy} \sum_{\bk, \bk^{\prime}} v^2_{X, \bk} 
e^\dagger_{X,\bk} e^\dagger_{X,-\bk}  
v^2_{Y, \bk^\prime}  e_{Y,-\bk^\prime} e_{Y,\bk^\prime} +h.c.
\ee
The set of Eq.s\ \pref{Hsc}-\pref{Hxy} is solved in the mean-field approximation 
by defining the orbital-dependent SC order parameters for the hole ($\D^{yz}_h, 
\D^{xz}_h$) and electron ($\D^{yz}_e, \D^{xz}_e, \D^{xy}_X,\Delta^{xy}_Y$) 
pockets. The self-consistent equations at $T=0$ reads: 
\bea
\lb{deltahyz}
\Delta^{yz}_h &=&-g_X \sum_\bk u^2_{X,\bk} \big(u^2_{X,\bk}\Delta^{yz}_e+v^2_{X,\bk}\Delta^{xy}_X\big)
/E_{X,\bk} \\
\lb{deltahxz}
\Delta^{xz}_h &=& -g_Y \sum_\bk  u^2_{Y,\bk} \big(u^2_{Y,\bk}\Delta^{xz}_e+v^2_{Y,\bk}\Delta^{xy}_Y\big)
/E_{Y,\bk}\\
\lb{deltaeyz}
\Delta^{yz}_e &=& -g_X \sum_\bk  u^2_{\G,\bk}
\big( u^2_{\G,\bk} \Delta^{yz}_h +v^2_{\G,\bk} \Delta^{xz}_h \big)/E_{\G,\bk} \\
\lb{deltaexz}
\Delta^{xz}_e &=& -g_Y \sum_\bk  v^2_{\G,\bk}
\big( u^2_{\G,\bk} \Delta^{yz}_h +v^2_{\G,\bk} \Delta^{xz}_h \big)/E_{\G,\bk},\\
\lb{deltaxxy}
\Delta^{xy}_X &=&- g_{xy} \sum_\bk  v^2_{Y,\bk}\big(u^2_{Y,\bk}\Delta^{xz}_e+v^2_{Y,\bk}\Delta^{xy}_Y\big)
/E_{Y,\bk}\\
\lb{deltayxy} 
\Delta^{xy}_Y &=& -g_{xy} \sum_\bk  v^2_{X,\bk}\big(u^2_{X,\bk}\Delta^{xz}_e+v^2_{X,\bk}\Delta^{xy}_X\big)
/E_{X,\bk}
\eea
Here $E_{\ell,\bk}=\sqrt{\e_{\ell,\bk}^2+\Delta_{\ell,\bk}^2}$ is the dispersion 
in the SC state, where $\e_{\ell,\bk}$ is the band dispersion on each pocket 
$\ell=\G,X,Y$ above $T_c$ and $\Delta_{\ell,\bk}$ is the band gap defined as: 
\bea 
\lb{deltag}
\Delta_{\G,\bk} &=&  u^2_{\G,\bk} \Delta^{yz}_h +v^2_{\G,\bk} \Delta^{xz}_h,\\
\lb{deltax}
\Delta_{X,\bk} &=& u^2_{X,\bk} \Delta^{yz}_e+v^2_{X,\bk} \Delta^{xy}_X,\\
\lb{deltay}
\Delta_{Y,\bk} &=& u^2_{Y,\bk} \Delta^{xz}_e +v^2_{Y,\bk} \Delta^{xy}_Y.
\eea

\subsection*{Superconducting Gaps Anisotropy}

The overall momentum dependence of the band gaps is determined by the 
interplay between the momentum dependence of the orbital factors and the 
hierarchy of the orbital SC order parameters. In the absence of 
nematic order Eq.s \pref{deltahyz}-\pref{deltay} preserve the symmetry in the 
exchange of the $xz/yz$ orbitals. Thus $\Delta^{xz}_h=\Delta^{yz}_h$ and the gap 
on the $\G$ pocket Eq.\ \pref{deltag} is constant, since 
$u^2_{\G,\bk}+v^2_{\G,\bk}=1$. In the nematic state the band structure breaks 
the $C_4$ symmetry, making $v^2_{\G,\bk}\gg u^2_{\G,\bk}$ (see Fig.\ 
2d), and also the SC orbital parameters $\Delta^{xz}_h$ and 
$\Delta^{yz}_h$ are in general different. However, as we shall see below, for 
isotropic pairing $g_X=g_Y$ the gaps anisotropy is the wrong one. The 
experimentally-observed anisotropy can only be achieved making $\Delta^h_{yz}\gg 
\Delta^h_{xz}$, that follows from the nematic pairing mechanism $g_X>g_Y$ 
provided by spin-nematic OSSF.

\begin{figure}[b]
\includegraphics[angle=0,width=0.94\linewidth]{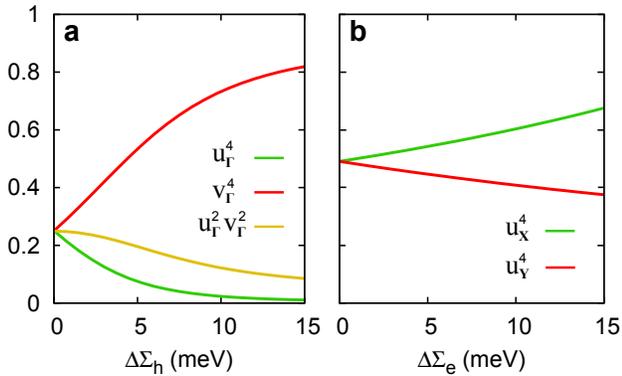}
\caption{Angular-averaged orbital-weight overlaps. ({\bf a},{\bf b}) 
Nematic-splitting dependence of the angular-averaged orbital-weight overlaps 
appearing in the SC gap equations Eq.s\ \pref{deltahyz}-\pref{deltaexz}. } 
\label{fig3}
\end{figure}
To understand the effect of the band-structure nematic reconstruction on the SC 
gap anisotropy we show in Fig.\ 3 the evolution of the orbital-factors 
overlaps appearing in Eq.s\ \pref{deltahyz}-\pref{deltaexz}, where we define the 
angular average of a given function as $\langle f(\bk)\rangle\equiv\int 
d\theta/(2\pi) f(k_F(\theta))$, with $k_F(\theta)$ FS wavevector of a given 
pocket. 
We can in first approximation neglect the pairing in the subleading $xy$ channel 
and consider only what happens in the $xz/yz$ orbital sector. As mentioned 
above, the nematic splitting on the electron pockets leads to a moderate 
enhancement of the $yz$ factor appearing in Eq.\ \pref{deltahyz} with respect to 
the $xz$ in Eq.\ \pref{deltahxz}, i.e $\langle u^4_X\rangle \gtrsim  \langle 
u^4_Y\rangle$, Fig.\ 3b. This effect, recently highlighted while 
discussing the $k_z=\pi$ FS cut\cite{Kang_PRL2018}, is however too small to 
account for the observed hole-gap anisotropy at $k_z=0$. In fact, the strong 
modification of the hole-pocket orbital factors implies that $\langle 
u^4_\G\rangle\ll \langle u^2_\G v^2_\G\rangle< \langle v^4_\G\rangle$, Fig.\ 3a. 
Thus, by neglecting logarithmic corrections in the gap ratios, from 
Eq.s \pref{deltahyz}-\pref{deltaexz} one obtains that 
\be
\lb{estim}
\frac{\Delta_e^{yz}}{\Delta_e^{xz}}\simeq \frac{g_X}{g_Y}\frac{ \langle u^2_\G v^2_\G 
\rangle}{\langle v^4_\G\rangle} \simeq 0.1 \frac{g_X}{g_Y}
\ee
and 
\be
\lb{estim_h}
\frac{\Delta_h^{yz}}{\Delta_h^{xz}}\simeq \frac{g_X}{g_Y}\frac{ \langle u^4_X \rangle}{\langle 
u^4_Y\rangle}\frac{ \Delta_e^{yz}}{\Delta_e^{xz}} \simeq 1.8 \frac{ g_X}{g_Y} \frac{ \Delta_e^{yz}}{\Delta_e^{xz}}.
\ee
Note that Eq.s\pref{estim}-\pref{estim_h} are almost unaffected once the 
$xy$ pairing channel is taken into account.
From Eq.s\pref{estim}-\pref{estim_h} it follows that an isotropic pairing 
interaction $g_X=g_Y$ (as considered in ref. \onlinecite{Kang_PRL2018}) would lead to a 
suppression of the $yz$ gap 
parameters. 
\begin{figure}[b]
\includegraphics[angle=0,width=0.94\linewidth]{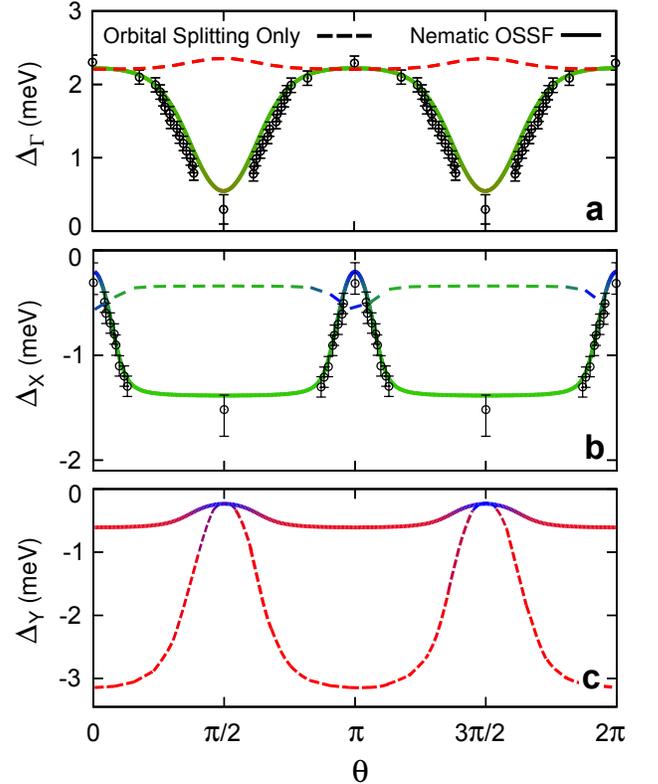} 
\caption{Angular dependence of the SC band gaps. ({\bf a}) SC gap on the hole , 
({\bf b}) on the $X$ electron pocket, ({\bf c}) on the $Y$ electron pocket. 
Dashed lines are the results for the isotropic pairing $g_X/g_Y=1$, while solid 
line for nematic pairing $g_X/g_Y\sim 21$. The color code accounts for the 
orbital component of the SC gap (green $yz$, red $xz$, blue $xy$), given by the 
product of the SC orbital parameter and the orbital weight, according to the 
definitions \pref{deltag}-\pref{deltay}. For comparison we reproduce the 
experimental gap values with standard deviations from ref.\ 
\onlinecite{Sprau_Science2017} (black circles).} 
\label{fig4}
\end{figure}
At the $\G$ pocket, where the $yz$ orbital character is also strongly suppressed 
by nematicity ($u_\G^2\ll v_\G^2$, Fig.\ 3a), the band gap would have 
only $xz$ character, $\Delta_{\G,\bk}\simeq \Delta_h^{xz}v^2_{\G,\bk}$, leading 
to a small modulation with a relative maximum at $\theta=\pi/2$ (dashed line in 
Fig.\ 3a), in contrast with the experimental findings. On the other 
hand, the OSSF-mediated anisotropic pairing with  $g_X/g_Y\gg1$ gives a 
substantial {\it enhancement} of the $\Delta_h^{yz}/\Delta_h^{xz}$ ratio. This 
leads to $\Delta_{\G,\bk}\simeq \Delta_h^{yz}u^2_{\G,\bk}$, in agreement with 
the band-gap anisotropy observed experimentally as shown in Fig.\ 4a, 
where the numerical solutions of Eq.s\ \pref{deltahyz}-\pref{deltaexz}  are 
reported along with the experimental data of\cite{Sprau_Science2017}. Here the 
colour code does not refer to the orbital content of the pocket, as in Fig.\ 1, 
but to the orbital content of the SC gap function, that is 
determined by the product of the SC order parameter times the orbital weight in 
each sector, Eq.s \pref{deltag}-\pref{deltay}. 

The anisotropy  $g_X/g_Y=21$ 
extracted from this analysis is rather large, since one needs to overcome the 
strong suppression of the $yz$ orbital due to nematic reconstruction at the hole 
pocket: one needs at least $g_X/g_Y\gtrsim 2$ (not shown) to start to see the 
correct symmetry of the gap at $\G$, i.e. a maximum at $\theta=0$. The value of 
$g_X/g_Y$ obtained by the SC-gaps analysis is compatible with the anisotropy of 
the OSSF used to reproduce the orbital selective shrinking of the FS 
in the nematic phase\cite{Fanfarillo_PRB2016} as discussed in Supplementary Note 3. 
In principle, the nematic-pairing 
anisotropy could also be estimated by the direct measurements of the SF. 
However, while it has been established that in the nematic phase SF are stronger 
at $(\pi,0)$ than at $(\pi,\pi)$\cite{Rahn_PRB2015, Wang_NatMat2016, 
Wang_NatCom2016}, the different intensity expected at $(\pi,0)$ and $(0,\pi)$ 
has not been measured yet in detwinned samples.

The gap obtained for the $X$ pocket is shown in Fig.\ 4b. Its value is 
also in overall in agreement with the STM experimental 
data\cite{Sprau_Science2017}. To reproduce the experimental value of the $xy$ 
component we needed a small ($|g_{xy}|\ll g_X$) attractive interband interaction 
between the two electron-like pockets. In fact, a negative $g_{xy}$ guarantees, 
from Eq.s \pref{deltaxxy}-\pref{deltayxy}, that the SC $xy$ order parameters on 
both electron pockets have the opposite sign with respect to the one at the hole 
pockets, as required by the dominant spin-mediated channel. In contrast, a 
repulsive $g_{xy}$ induces a frustration that turns out in a gap with nodes 
along the Fermi surface\cite{Kemper_NJP2010}. Even though this has been recently 
suggested by specific-heat measurements\cite{Hardy_arxiv2018}, the STM 
data\cite{Sprau_Science2017} shown for comparison exclude the presence of nodes 
and force us to consider a negative $g_{xy}$. It is important to stress that, 
even though the full set of equations \pref{deltahyz}-\pref{deltayxy} must be 
solved self-consistently, adding or not the $xy$ channel is not relevant for 
what concerns the understanding of the gap behavior in the $xz/yz$ sector, 
especially for the gap anisotropy at the $\G$ pocket. 
For the sake of completeness we report in Fig.\ 4c also the gap on the 
$Y$ pocket, that has not been resolved so far in STM\cite{Sprau_Science2017}. As 
one can see, for the electronic pockets an isotropic pairing   $g_X=g_Y$ would 
lead to a strong difference between the absolute gap values at $X$ and $Y$, due 
to the effect of nematic ordering at the electronic pockets, as one understands 
from Eq.\ \pref{estim} above. In contrast nematic pairing leads to more similar 
gap values, which can be hardly disentangled experimentally, explaining why  
recent ARPES results claiming to resolve the $Y$ pocket do not report 
appreciable significant gap differences on the two electron pockets 
\cite{Kushirenko_PRB2018}. The differences between the $X$ and $Y$ gaps due to 
the nematic pairing could however have implications for the thermal probes 
sensibles to single-particle excitations. We leave the analysis of those effects 
for future work.

\begin{figure}[b] \includegraphics[angle=0,width=0.99\linewidth]{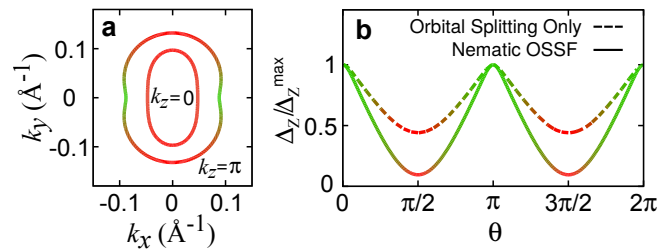} 
\caption{FS and SC gap for the hole pocket at $k_z=\pi$. ({\bf a}) FS at $k_z=0$ 
and $k_z=\pi$ in the nematic phase. At $k_z=\pi$ the hole pocket retains a full 
$yz$ orbital character at $\theta =0$. ({\bf b}) Angular dependence of the SC 
gap $\Delta_Z$ renormalized to its maximum value obtained using the same 
$g_X/g_Y$ extracted from the $k_z=0$ gap fit (see Supplementary Note 3). Same color code of 
Fig.\ 4. The $\Delta_Z$ maximum at $\theta=0$ is obtained already for 
isotropic pairing  $g_X=g_Y$ (dashed lines), see also\cite{Kang_PRL2018}. The 
nematic pairing (solid lines) further enhances the gap anisotropy, leading to 
larger relative variations on the $Z$ pocket, in agreement with ARPES 
experiments\cite{Kushirenko_PRB2018}} 
\label{fig5}
\end{figure}
Recently, the $k_z$-dependence of the gap anisotropy on the hole pocket has been 
investigated\cite{Kushirenko_PRB2018}, and it has been shown that the 
$\Delta_\G(\theta=0)/\Delta_\G(\theta=\pi/2)$ anisotropy increases as one moves 
from the  $k_z=0$ to the $k_z=\pi$ cut. Even though we did not consider a full 
3D model, this effect can be understood by considering the variations of the 
hole-pocket orbital content when moving from $k_z=0$ to $k_z=\pi$ (Z point). The 
larger size of the hole pocket at $Z$ makes its orbital content less sensitive 
to nematic ordering and spin-orbit mixing, so that it still preserves a marked 
$yz$ character around $\theta=0$ (ref.s \onlinecite{Xu_PRL2016, 
Hashimoto_NatCom2018, Kang_PRL2018}), with $u_\G\sim \cos\theta$ and $v_\G\sim 
\sin\theta$ also in the nematic phase (Fig. 5a).  In this situation 
$\langle u^4_\G\rangle \sim \langle v^4_\G\rangle $ so that the enhancement 
$\langle u^4_X\rangle>\langle u^4_Y\rangle $ of the orbital factors in the 
electron pockets is enough to guarantee that $\Delta_h^{yz}>\Delta_h^{xz}$, 
leading to a hole-pocket gap anisotropy compatible with the measurements even 
when $g_X=g_Y$, as recently shown in\cite{Kang_PRL2018} (dashed line Fig. 5b). 
On the other hand, by retaining the same ratio $g_X/g_Y$ extracted 
from the  $k_z=0$ gap fit (solid line Fig. 5b),  we find an increase of 
the anisotropy when moving from the $\G$ to the $Z$ pocket. While this is 
consistent with the observations in pure\cite{Kushirenko_PRB2018} and 
S-doped\cite{Xu_PRL2016} FeSe, other groups\cite{Hashimoto_NatCom2018, 
Rhodes_arxiv2018} report instead an overall smaller gap at $k_z=0$. The analysis 
of SC fluctuations above $T_c$, could provide an alternative experimental test 
to clarify the 3D behavior. As shown in\cite{Fanfarillo_SupSciTech2014}, the 
crossover from 2D to 3D character of the fluctuation contribution to the 
paraconductivity is controlled by the $k_z$ dependence of the pairing 
interactions. This effect, used to explain the measurements in 122 
systems\cite{Rullier_PRL2012}, could be tested in FeSe as well.

\section*{DISCUSSION}
The $C_4$ symmetry breaking of paramagnetic SF is a consequence of SF interactions 
beyond RPA\cite{Fernandes_NatPhys2014, Fanfarillo_PRB2018}. As a consequence 
the effects of the nematic SF pairing $g_X>g_Y$ highlighted in the present 
work cannot be captured by microscopic models where the SF are described at RPA 
level, even when RPA fluctuations are computed using the nematic reconstruction 
of the band structure\cite{Kreisel_PRB2017,Kreisel_arxiv2018}. An alternative route 
followed in\cite{Kreisel_PRB2017, Sprau_Science2017} amounts to start from band 
dispersions fitted to ARPES data and to account phenomenologically for the role 
of correlations. 
The so-called orbital differentiation of the electronic mass 
renormalization due to local electronic interactions has been studied in 
DMFT-like calculations in the tetragonal 
phase\cite{deMedici_PRB2005,deMedici_PRL2014}, which found in particular a 
larger renormalization of the $xy$ orbital with respect to the $xz/yz$ ones. In 
addition, correlations can also cooperate to enhance the $xz/yz$ orbital 
differentiation induced by other nematic mechanisms\cite{Fanfarillo_PRB2017}. 
Inspired by these results, the authors of ref.s\ \onlinecite{Kreisel_PRB2017, 
Sprau_Science2017} added phenomenologically orbital-dependent quasiparticle 
spectral weights, $Z_{orb}$, in the RPA-based calculation of the pairing interaction. By 
using $Z_{xy}\ll Z_{xz}<Z_{yz}$ they obtain the twofold result to make the $Y$ 
pocket incoherent, explaining why it does not show up in the STM 
analysis\cite{Sprau_Science2017,Kostin_NatMat2018}, and to move the maximum of 
SF from  $\bQ=(\pi,\pi)$ to $\bQ=\bQ_X$ \cite{Kreisel_arxiv2018}, explaining the 
neutron-scattering experiments\cite{Wang_NatMat2016,Wang_NatCom2016} and the 
observed gap hierarchy. However, this approach presents some inconsistencies. One 
issue is methodological: by using independent parameters to renormalize the band 
structure (that is fitted from the experiments) and to define the residua of the 
Green's functions, one misses the strict relation between these two quantities.  
On the other hand, by implementing this relation self-consistently, as done for 
example in\cite{Hu_arxiv2018}, it is not obvious how one can reconcile the large 
Fermi-velocity anisotropy implicit in the $Z_{xz}<Z_{yz}$ relation with the 
experimental band structure, that is well reproduced accounting only for a 
crystal-field splitting of the tetragonal band structure having $Z_{xz}=Z_{yz}$ 
\cite{Coldea_Review2018, Liu_PRX2018, Rhodes_arxiv2018}. A second issue arises 
by the comparison with experiments. The route followed in \cite{Kreisel_PRB2017, 
Sprau_Science2017} is equivalent to rewrite the SC gap e.g. on the $\G$ pocket 
as: 
\be 
\lb{deltagZ}
\Delta_{\G,\bk}=  Z_{yz}u^2_{\G,\bk} \Delta^{yz}_h +Z_{xz}v^2_{\G,\bk} \Delta^{xz}_h
\ee
In our case, Eq.\ \pref{deltag},  the predominance of the SC $yz$ orbital 
component is achieved via $\Delta_{yz}^h\gg \Delta^h_{xz}$, as guaranteed by the 
nematic-pairing condition $g_X\gg g_Y$. Instead in Eq.\ \pref{deltagZ} this is 
due mainly to the rescaling of the orbital occupation factors  by the 
corresponding spectral weights $Z_{yz/xz}$. By assuming $Z_{yz}\gg Z_{xz}$
\cite{Kreisel_PRB2017, Sprau_Science2017} one finds $\Delta_{\G,\bk}\simeq  
Z_{yz}u^2_{\G,\bk} \Delta^{yz}_h $, consistently with the measured gap 
anisotropy. However, the rescaling of the  $yz$ orbital occupation to 
$Z_{yz}u^2_{\G,\bk}$ is operative not only on the SC gap function, but also on 
the band structure above $T_c$. This  restores the  $yz$ character of the $\G$ 
pocket\cite{Kreisel_arxiv2018}, in contrast with ARPES measurements which 
clearly indicate\cite{Liu_PRX2018, Rhodes_arxiv2018} its predominant $xz$ 
character. 

To reconcile ARPES with RPA-based calculations of the spin-mediated pairing 
interactions the authors of\cite{Rhodes_arxiv2018} use the alternative approach 
to remove intentionally the contribution of the $Y$ pocket from the RPA-mediated 
pairing interaction. This is equivalent to put $g_Y=0$ in Eq.s\ 
\pref{deltahyz}-\pref{deltaexz}, so that $\Delta_h^{xz}=0$ and the modulation of 
the gap at $\Gamma$ follows again the $yz$ orbital weight, even if it is largely 
subdominant. With respect to these approaches, the main advantage of our model 
is to provide, via the orbital selectivity of the OSSF, a mechanism able to 
achieve the $g_X>g_Y$ nematic pairing without  affecting strongly the 
quasiparticle spectral weights, while a main disadvantage is the lack of a 
theoretical justification for the missing $Y$ pocket. However, we cannot help 
noticing that this point is also controversial from the experimental point of 
view, due to different reports claiming to observe \cite{Fanfarillo_PRB2016, 
Kushirenko_PRB2018} or not\cite{Sprau_Science2017, Rhodes_arxiv2018} the Y 
pocket.

In summary, our work provides a paradigm for the emergence of superconductivity
in FeSe from an orbital-selective nematic SF mechanism. By
combining the orbital ordering induced by the nematic shrinking of the Fermi
surface pockets below the nematic transition with the anisotropic pairing interaction
mediated by nematic SF, we explain the  gap hierarchy reported
experimentally on hole and electron pockets, and its variation with $k_z$. Our
findings offer also a fresh perspective on previous attempts to explain the SC
properties of FeSe, highlighting from one side the crucial role of spin-mediated
pairing, and from the other side clarifying the importance of spin-spin
interactions beyond RPA level. This result represents then a serious challenge
for a full microscopic approach, that must account self-consistently for the
emergence of Ising-nematic SF below the nematic transition
temperature. 

\section*{METHODS}

\subsection*{\bf Pairing by Orbital selective spin fluctuations}

The mean-field equations for the pairing Hamiltonian, Eq.s \pref{Hsc}-\pref{Hxy}, can be
easily derived by defining the orbital-dependent SC order 
parameters for the hole ($\D^{yz}_h, \D^{xz}_h$) and electron ($\D^{yz}_e, 
\D^{xz}_e$) pockets as: 
\bea
\Delta^{yz}_e &=& -g_X  \langle u^2_{\G, \bk} h_\bk h_{-\bk}\rangle,\\
\Delta^{xz}_e &=& -g_Y  \langle v^2_{\G, \bk} h_\bk h_{-\bk}\rangle, \\
\Delta^{yz}_h &=& -g_X  \langle u^2_{X, \bk} e_{X,\bk} e_{X,-\bk}\rangle, \\
\Delta^{xz}_h &=&- g_Y  \langle u^2_{Y, \bk} e_{Y,\bk} e_{Y,-\bk}\rangle,\\
\Delta^{xy}_X &=& -g_{xy}  \langle v^2_{Y, \bk}e_{Y,\bk} e_{Y,-\bk} \rangle,\\
\Delta^{xy}_Y &=& -g_{xy}  \langle v^2_{x, \bk}e_{X,\bk} e_{X,-\bk} \rangle,
\eea

The corresponding self-consistent BCS equations at $T=0$
are the ones reported in the text, Eq.s\ \pref{deltahyz}-\pref{deltayxy}. To 
solve them we introduce polar coordinates and we approximate the orbital factors 
and the density of states with their values at the Fermi level for each pocket. 
This implies that  the various integrals can be computed as for example:
\bea
&\sum_\bk& u^2_{X,\bk} \ \frac{\Delta_{X, \bk}}{E_{X,\bk}}
=\int \frac{kdkd\theta }{(2\pi)^2} u^2_{X}(\theta)
\frac{\Delta_{X}(\theta)}{\sqrt{\e_{X,\bk}^2+\Delta^2_X(\theta)}} \nn\\
&&=\int \frac{d\e d\theta }{(2\pi)}N_X(\e_F,\theta)
u^2_{X}(\theta)\frac{\Delta_{X}(\theta)}{\sqrt{\e^2+\Delta^2_X(\theta)}} \nn \\
&&=\int \frac{d\e d\theta }{(2\pi)}N_X(\e_F,\theta)
u^2_{X}(\theta)\Delta_{X}(\theta)\log\frac{\omega_D}{\Delta_{X}(\theta)}
\lb{integral}
\eea
where we defined $u^2_{X}(\theta)\equiv u^2_{X}(k_F(\theta))$ and 
$\Delta_X(\theta)\equiv \Delta_e^{yz}u^2_{X}(\theta) + \Delta_X^{xy}v^2_{X}(\theta)$. 
The cut-off $\omega_D$ 
represents the range of the spin-mediated pairing interaction, and it has been 
taken of order of 0.1 eV. The angular dependent density of state is defined as 
usual as $N_X(\e_F,\theta)=\int 
(kdk)/(2\pi)\delta(\epsilon_F-\epsilon_{X,\bk})=k_F(\theta)/2\pi |v_F(\theta)|$, 
where $k_F(\theta)$ and $v_F(\theta)$ are the wavevector and velocity at the 
Fermi level, respectively. For a parabolic band dispersion $N_X(\e_F,\theta)$ 
reduces to an angular-independent constant, as usual. Even though in Eq.\ 
\pref{integral} the angular integration involves both the orbital factor and the 
density of states, we checked that the results do not change considerably if the 
angular-averaged density of states is taken outside the integral. For this 
reason, accounting separately for the angular averages of the orbital factors 
alone, as shown in Fig.\ 3, allows one to have a rough estimate of the 
numerical results, as discussed in the text. The results of the full numerical 
self-consistent calculations of Eq.s\ \pref{deltahyz}-\pref{deltaexz} are 
displayed in Fig.\ 4-5 for $g_X/g_Y=21$ and $|g_{xy}|/g_X = 
0.076$. The numerical values of the band parameters can be found in 
Supplementary Note 3. 

\section*{ACKNOWLEDGEMENTS}
We acknowledge M. Capone, A. Chubukov and P. Hirschfeld 
for useful discussions, and M. Capone  for critical reading of the manuscript. 

\section*{COMPETING INTERESTS}
The authors declare no competing interests.

\section*{AUTHOR CONTRIBUTIONS}
L.F. conceived the project with inputs from all coauthors. L.F. and L.B.
performed the numerical calculations. All the authors contributed to the data
analysis, to the interpretation of the theoretical results and to the writing of
the text.

\section*{FUNDING}
We acknowledge financial support  by Italian MAECI under the collaborative 
Italia-India project SuperTop-PGR04879, by MINECO (Spain) via Grants 
No.FIS2014-53219-P and by Fundaci\'on Ramon Areces. We acknowledge the cost 
action Nanocohybri CA16218. 

\section*{DATA AVAILABILITY}
The authors declare that the data supporting the findings of this study are
available within the paper and its supplementary notes.

\section*{SUPPLEMENTARY MATERIAL}

\section*{Supplementary Note 1: Band structure}
To describe the band structure of FeSe above the superconducting (SC) 
transition, we adapt the orbital model of ref.\ \onlinecite{Cvetkovic_PRB2013}. The 
effective band-mass parameters are extracted from ARPES measurements. These 
values, considerably smaller than the ones predicted by LDA, are usually 
reproduced remarkably well by DMFT-based calculations \cite{deMedici_PRL2014}. 
However, both LDA and DMFT fail in the description of the measured Fermi 
surface (FS), that are always smaller than expected. Such a FS shrinking 
\cite{Ortenzi_PRL2009}, present already well above the nematic transition 
\cite{Fanfarillo_PRB2016, Kushnirenko_PRB2017}, and the nematic splitting, can be 
explained instead within our low-energy approach by accounting for the orbital 
selective spin fluctuations (OSSF). As detailed in ref.\ \onlinecite{Fanfarillo_PRB2016}, 
the orbital-dependent self-energy corrections due to the exchange of spin 
fluctuations at $\bQ_X$ and $\bQ_Y$ lead in general to a temperature-dependent 
FS shrinking. Due to the anisotropy of the OSSF in a spin-nematic transition, 
this results in the nematic orbital splitting below $T_s$. 

To illustrate the model, let us start from the low-energy Hamiltonian around the 
$\Gamma$ point in the orbital space: 
\be
\lb{h0}
H^\G=\sum_{\bk,\s} \psi^{\G,\dagger}_{\bk\s} \hat H_{0\bk}^{\G} \psi^\G_{\bk\s}.
\ee
Here the spinor is defined as 
$\psi^{\Gamma}_{\bk,\s}=(c_{yz,\bk,\s},c_{xz,\bk,\s})$. Taking into account the 
real part of the isotropic $\Sigma_0^\G$ and anisotropic $\Delta\Sigma_h$ 
components of the self-energy, responsible for the nematic shrinking, one has 
that: 
\be
\lb{Hin_nem}
\hat H^\G = \begin{pmatrix}
h_{0\bk}^{\G} + \Sigma^\G_{0} + h_{3\bk}^\G - \Delta\Sigma_h \ \ & \ \ h_{1\bk}^{\G} \\
 h_{1\bk}^\G \ \ & \ \ h_{0\bk}^\G + \Sigma^\G_0 - h_{3\bk}^l +\Delta\Sigma_h \\
\end{pmatrix}
\ee
where the $h^\G_{i,\bk}$ components read
\bea
h_{0,\bk}^{\G} &=& \epsilon_\G -a_\G \bk^2, \nn \\
h_{1,\bk}^{\G} &=& -2 b_\G k_x k_y, \nn \\
\lb{hG}
h_{3,\bk}^{\G} &=& b_\G (k_x^2 -k_y^2),
\eea
and we defined $\Sigma_0^\G\equiv (\Sigma^\G_{xz}+\Sigma^\G_{yz})/2$ and 
$\D\Sigma_h\equiv (\Sigma^\G_{xz}-\Sigma^\G_{yz})/2$. As discussed in ref.s\ 
\onlinecite{Ortenzi_PRL2009, Fanfarillo_PRB2016} the self-energy corrections are negative at 
the hole pocket, so the $\Sigma_0^\G$ term accounts in general for an uniform 
shrinking of the pocket, present already in the paramagnetic phase. Below the 
structural transition temperature, $T_S$, nematic spin fluctuations induce 
larger corrections in the $\bQ_X$ direction, which translate in a 
$|\Sigma^\G_{yz}|>|\Sigma^\G_{xz}|$, so that $\Delta\Sigma_h>0$ (ref.\onlinecite{Fanfarillo_PRB2016}). 
Taking into account also the spin-orbit splitting it is 
easy to see\cite{Cvetkovic_PRB2013} that the Hamiltonian \pref{Hin_nem} gets an 
additional term $\pm\l/2\hat\sigma_2$ in the $\pm$ spin sector. As a consequence 
the eigenvalues defining the bands are given by: 
\be
\e_{\G,\bk,\pm}=h_{0,\bk}^\G-|\Sigma_0^\G|\pm\sqrt{{h_{1,\bk}^\G}^2+(h_{3,\bk}^\G-\Delta\Sigma_h)^2 +(\l/2)^2},
\ee
where the $+/-$ refers to the outer/inner pocket, respectively. Since the two 
spin sectors have the same energy dispersion we drop from now on any explicit 
dependence on the spin index. By introducing the orbital weights: 
\bea
u_{\G,\bk}^2 &=&\frac{1}{2} \Bigg( 1+\frac{h_{3,\bk}^\G 
-\Delta\Sigma_h}{\sqrt{{h^\G_{1,\bk}}^2+ (h^\G_{3,\bk} -\Delta \Sigma_h)^2 +(\l/2)^2}}\, \Bigg) \nn  \\
v_{\G,\bk}^2 &=&\frac{1}{2} \Bigg( 1-\frac{h_{3,\bk}^\G
- \Delta\Sigma_h}{\sqrt{{h^\G_{1,\bk}}^2+ (h^\G_{3,\bk} -\Delta \Sigma_h)^2 +(\l/2)^2}}\, \Bigg)\nn\\
\lb{uvG}
\eea
one can also define the rotation from the orbital to the band basis
\be
\lb{hole_fer}
\begin{pmatrix}
 h_{+,\bk} \\
 h_{-,\bk} \\
\end{pmatrix} = 
\begin{pmatrix}
u_{\Gamma,\bk} &-v_{\Gamma,\bk}\\
v_{\Gamma,\bk} & u_{\Gamma,\bk}\\
\end{pmatrix}
\begin{pmatrix}
 c_{yz,\bk} \\
 c_{xz,\bk} \\
\end{pmatrix} 
\ee
where $h^\dagger_+/h^\dagger_-$ is the creation operator of a quasiparticle in 
the outer/inner pocket, respectively. Since in FeSe only the outer pocket 
crosses the Fermi level, throughout the main text we dropped the $+$ index and 
we simply referred to $h_\bk$ and $\e_{\G,\bk}$ as the fermionic operator and 
bare dispersion of the outer hole pocket. In the paramagnetic phase 
($\D\Sigma_h=0$) and in the absence of spin-orbit interaction the two hole bands 
have a simple parabolic dispersion $\e_{\G,\bk,\pm}=\e_\G-|\Sigma_0^\G|-(a_\G\mp 
b_\G)\bk^2$. In this case the orbital weights only depend on the azimuthal angle 
$\theta$ measured with respect to $k_x=0$, so that $u_{\G,\bk}=\cos\theta$ and 
$v_{\G,\bk}=\sin \theta$. However, the spin-orbit interaction and the nematic 
splitting mix the two orbital characters, leading to the angular dependence of 
the orbital weights at the Fermi level shown in Fig. 2d of the main text.

For the $X/Y$ pockets the general structure is analogous to Eq.\ \pref{h0},
provided that the spinors are now defined as $\psi^{X}_\bk=(c_{yz,\bk},c_{xy,bk})$ and
$\psi^{Y}_\bk=(c_{xz,\bk},c_{xy,\bk})$. In addition, since the $xy$ orbital is not
affected by OSSF, one has in general 
\be
\lb{HX_nem}
\hat H^X_\bk = \begin{pmatrix}
h_{0,\bk}^X+ \Sigma^X + h_{3,\bk}^X\ \ & \ \ -ih_{2,\bk}^X \\
 ih_{2,\bk}^X \ \ & \ \ h_{0,\bk}^X-h_{3,\bk}^X \\
\end{pmatrix}
\ee
for the $X$ pocket, with
\bea
h_{0,\bk}^{X} &=& (h_{yz,\bk}+h_{xy,\bk})/2 \nn \\
h_{2,\bk}^{X} &=& v k_y \nn\\
\lb{hX}
h_{3,\bk}^{X} &=& (h_{yz,\bk}-h_{xy,\bk})/2 - b (k_x^2 -k_y^2) 
\eea
where $h_{yz,\bk} = -\epsilon_{yz} + a_{yz} \bk^2$ and $h_{xy,\bk} = - \epsilon_{xy} + 
a_{xy} \bk^2$. Ana\-lo\-gous expressions hold for the $Y$ pocket provided that 
one exchange the role of $k_x$ and $k_y$, $h_{i_\bk}^Y(k_x,k_y)=h_{i,\bk}^X(k_y,k_x)$, and 
$\Sigma_{yz}^X$ is replaced by $\Sigma_{xz}^Y$. At the electron pockets the 
self-energy corrections are positive, so that the $\Sigma_{yz/xz}^{X/Y}$ terms 
lead again to an upward shift of the $yz/xz$ orbitals. In the spin-nematic state 
$\Sigma_{yz}^X>\Sigma_{xz}^Y$ so the $yz$ sector of the $X$ pocket shrinks 
further and the $xz$ part of the $Y$ pocket expands, see Fig. 1 in the main 
text. The nematic order parameter at the electron pockets is then defined as $\D 
\Sigma_e=(\Sigma_{yz}^X-\Sigma_{xz}^Y)/2>0$. Notice that in our approach the 
change of sign of the nematic splitting at the $\G$ and $M=(X,Y)$ point is a 
natural consequence of the self-energy corrections induced by nematic OSSF, as 
explained in \cite{Fanfarillo_PRB2016}. The $X/Y$ band dispersions are given by 
\be
\e^{X/Y}_{\bk,\pm}=h_{0,\bk}^{X/Y}+\Sigma^X/2\pm\sqrt{{h_{2,\bk}^X}^2+(h_{3,\bk}^X-\Sigma_{yz}^X/2)^2},
\ee
such that $\e^{X/Y}_{\bk,+}$ is the electronic band crossing the Fermi level at the $X/Y$ point. 
The rotation from the orbital to the band basis is defined now as
\be
\lb{ex_fer}
\begin{pmatrix}
 e_{X,\bk,+} \\
 e_{X,\bk,-} \\
\end{pmatrix} = 
\begin{pmatrix}
u_{X,\bk} &-iv_{X,\bk}\\
iv_{X,\bk} & u_{X,\bk}\\
\end{pmatrix}
\begin{pmatrix}
 c_{yz,\bk} \\
 c_{xy,\bk} \\
\end{pmatrix} 
\ee
with the orbital weights given by
\bea
u_{X,\bk}^2&=&\frac{1}{2}\bigg(1+\frac{h_{3,\bk}^X-\Sigma_{yz}^X/2}{\sqrt{{h_{2,\bk}^X}
^2+(h_{3,\bk}^X-\Sigma_{yz}^X/2)^2}}\bigg),\nn \\
\lb{uvX}
v_{X,\bk}^2&=&\frac{1}{2}\bigg(1-\frac{h_{3,\bk}^X-\Sigma_{yz}^X/2}{\sqrt{{h_{2,\bk}^X}
^2+(h_{3,\bk}^X-\Sigma_{yz}^X/2)^2}}\bigg),
\eea
At the $Y$ points the definitions are again equivalent, provided that one 
replaces $\Sigma_{yz}^X$ with $\Sigma_{xz}^Y$ and $k_x$ with $k_y$. As for the 
hole pocket, we drop the $+$ index and we refer to $e_{X/Y,\bk}$ and 
$\e_{X/Y,\bk}$ as the fermionic operators and energy dispersions of the 
electronic $X/Y$ pockets. In summary, the notations used in the main text are: 
\bea
\e_{\G,\bk,+}&\ra& \e_{\G,\bk} \quad h_{+,\bk}\ra h_{\bk} \quad \mathrm{hole \,\, pocket}\nn\\
\e_{X,\bk,+}&\ra& \e_{X,\bk} \quad e_{X,\bk,+}\ra e_{X,\bk} \quad X \mathrm{\,\, pocket}\nn\\
\e_{Y,\bk,+}&\ra& \e_{Y,\bk} \quad e_{Y,\bk,+}\ra e_{Y,\bk} \quad Y \mathrm{\,\, pocket}\nn
\eea
With these definitions in mind the rotation from the orbital to the band 
basis defined in Eq.s (1)-(3) of the main text are equivalent to Eq.\ 
\pref{hole_fer} and \pref{ex_fer} above.  

Finally, we notice that the present low-energy model describes 
properly the orbital character of the bands up to energy scale of order of 0.5 
eV around the Fermi level, beyond which additional $d$ orbitals should be taken 
into account \cite{Fernandes_Review2017}. Since both the nematic and SC 
transition involved much smaller energy scales, the results obtained within the 
present low-energy approach are expected to be robust with respect to the 
band-structure description obtained within  more sophisticated five- or 
ten-orbital models.

\section*{Supplementary Note 2: Orbital Selective Spin Fluctuations Model}
Once established the orbital composition of the low-energy model, one can
project the general interacting Hamiltonian including the  Hubbard and Hund
terms into the low-energy states. As shown in \cite{Fanfarillo_PRB2015,
Fanfarillo_PRB2018} one obtains that the effective low-energy interacting terms can
be written as 
\be
\lb{hint}
H_{int}= - \frac{\tilde{U}}{2} \sum_{\bq} \bS^{yz/xz}_{X/Y} \cdot \bS^{yz/xz}_{X/Y}.
\ee
Here $\tilde{U}$ is the intraorbital interaction renormalized at low energy and 
the intraorbital spin operators connecting hole and electron pocket are given by 
Eq.s (4)-(5) of the main text, that we rewrite here explicitly including also 
the contribution of the inner pocket, when present:
\bea
\lb{sx}
\bS_{X}^{yz} &=& \sum_\bk (u_{\G}h_{+}^{\dagger} 
+ v_{\G} h_{-}^{\dagger} )\, \vec{\s} \, u_{X} e_X, \\
\lb{sy}
\bS_{Y}^{xz} &=& \sum_\bk (-v_{\G}h_{+}^{\dagger} 
+ u_{\G} h_{-}^{\dagger} )\, \vec{\s} \, u_{Y} e_Y, 
\eea
where momentum dependence has been dropped for simplicity. 
The low-energy interacting Hamiltonian in Eq.\pref{hint} defines the OSSF: at 
low energy the hole pockets at $\G$ and the $X/Y$ electron pockets share only 
one orbital, the $yz$/$xz$ respectively. Thus the spin interactions along $x$ 
and $y$ has a single orbital character (see Fig 1 in the main text):
\bea
\langle \bS\cdot \bS\rangle (\bQ_X) &\Ra& \langle \bS^{yz}_{X}\cdot \bS^{yz}_{X}\rangle \nn \\
\langle \bS\cdot \bS\rangle (\bQ_Y) &\Ra& \langle \bS^{xz}_{Y}\cdot \bS_{Y}^{xz}\rangle   
\lb{corry}
\eea
By computing self-energy corrections of the orbital states coming from these 
OSSF one obtains orbital-dependent self-energy corrections, as shown in Eq.\ 
\pref{Hin_nem} and \pref{HX_nem} above. In addition, within a spin-nematic 
scenario the anisotropy of the spin fluctuations at different $\bQ$ vectors 
translates in the nematic splitting $\Delta\Sigma_h$, $\Delta\Sigma_e$ of the 
orbitals discussed previously. Here we argue that OSSF can also mediate an 
orbital-selective nematic pairing. The pairing model mediated by OSSF can be 
easily derived by rewriting the spin-spin interaction terms \pref{hint} in the 
pairing channel, using the projection of the orbital spin operator on the band 
basis encoded in Eq.s\ \pref{sx}-\pref{sy} above. The resulting pairing 
interaction is given by Eq. (8) of the main text.  

\section*{Supplementary Note 3: Model parameters for FeSe}
We solve self-consistently the set of BCS equations for realistic parameters for
the FeSe system in the nematic phase. 

Although the physical outcome of this work does not crucially depend on this, 
instead of using exactly the band parameters of ref. \onlinecite{Fanfarillo_PRB2016}, we 
will adjust them to fit a smaller value of the nematic splitting of the electron 
pockets reported afterwards in the literature, $\Delta\Sigma_e\simeq 15$ meV 
\cite{Fedorov_SciRep2016,Watson_NJP2017,Coldea_Review2018}.  
When computing self-consistently the spectral function of 
electrons coupled to spin modes in ref.\ \onlinecite{Fanfarillo_PRB2016}, we included the 
full frequency-dependent self-energies, thus we also effectively included the 
quasiparticle weight $Z_{spin}$ due to the orbital-dependent mass 
renormalization. However, since we checked that $Z_{spin}$ was of order one for 
the various orbitals, to reduce the number of parameters we decided in the 
present work to choose directly the band parameters which reproduce the 
experimental dispersions. This explains  the small quantitative differences 
between the values listed in Table \ref{bp} and those listed in ref.\ 
\onlinecite{Fanfarillo_PRB2016}. The list of the band parameters appearing in Eq.s 
\pref{hG}, \pref{hX} and used in the calculations are given in Table \ref{bp}. 
The spin-orbit coupling is assumed $\lambda$= 20 meV. We use 
$|\Sigma_{yz/xz}^\G| = 70/40$ meV and $\Sigma_{yz/xz}^{X/Y} = 45/15$ meV, in 
order to have $\Delta\Sigma_{h/e} = 15$ meV.
\begin{table}[tbh]
\begin{center}
\begin{tabular}{ccccccccccccccccccc}
\hline 
&              &    $\Gamma$   &   &\qquad \qquad \qquad&                       &X&                     &\\
\hline \hline
&$\epsilon_\G$ &               & 46&                    &$\epsilon_{xy}$ \ \ 72 & &$\epsilon_{yz}$\ \ 55&\\
\hline
& $a_\G$       &               &263&                    & $a_{xy}$\ \ 93        & &$a_{yz}$\ \ 101       &\\
\hline
& $b_\G$       &               &182&                     & $b$ \ \ \ 154          & &                     &\\
\hline
&              &               &   &                    & $v$ \ \ \ 144          & &                     & \\
\hline
\hline
\end{tabular}
\caption{Low-energy model parameters for FeSe in the nematic phase at $k_z=0$. 
All the parameters are in meV, the $k$ vector is measured in units $1/a \sim 
0.375$ \AA, where $a=a_{FeFe}$ is the lattice constant of the 1-Fe unit cell (so 
that $\tilde a=\sqrt{2}a=3.77$ \AA \, is the lattice constant of the 2Fe unit 
cell).} 
\label{bp}
\end{center}
\end{table} 

The $u$, $v$, factors defined by Eq.s \pref{uvG}, \pref{uvX}, computed using the 
above set of parameters, are shown in Fig. 2 of the main text. We reproduce the 
FS and their orbital distribution as experimentally observed by ARPES at $k_z=0$ 
(ref.s \onlinecite{Fanfarillo_PRB2016, Fedorov_SciRep2016, Watson_NJP2017, Coldea_Review2018, Rhodes_arxiv2018}), 
with the hole pocket having $k_F^x=0.056$ \AA$^{-1}$ and 
$k_F^y=0.11$ \AA$^{-1}$, the $X$ one $k_F^{x}=0.20$ \AA$^{-1}$ and 
$k_F^{y}=0.051$ and the $Y$, $k_F^{x}=0.10$ and $k_F^{y}=0.20$ \AA$^{-1}$ in the 
nematic phase.

\begin{figure}[h]
\includegraphics[angle=0,width=0.85\linewidth]{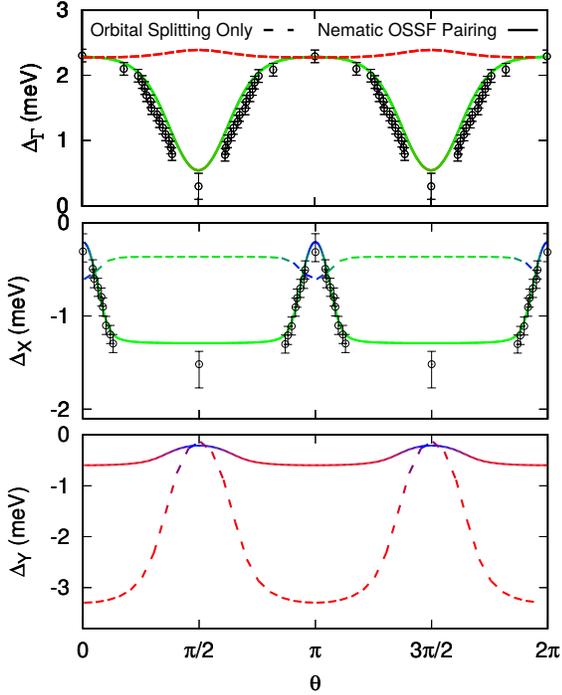} \caption{Numerical 
results for the band gaps on the $\Gamma$ (top), $X$ (middle) and $Y$ (bottom) 
pocket obtained considering a larger nematic 
splitting at $M$, $\Delta\Sigma_{e} = 20$ meV.} 
\label{figlarge}
\end{figure}

Using now the $T=0$ BCS equations, Eq.s (10)-(15) of the main text, we can fit 
the experimental data of ref.\ \onlinecite {Sprau_Science2017} using the SC couplings as 
fitting parameters. The results for the orbital SC order parameters,  Eq.s 
(10)-(15), are listed in Table \ref{orb_gap} while in Fig. 4 of the main text we 
showed the band gaps, Eq.s (16)-(18) of the main text. Notice that assuming 
isotropic pairing interactions $g_X=g_Y=1.03$ eV and accounting only for the 
orbital splitting effects, encoded in the $u,v$ factors, the SC gap at $\G$ 
presents opposite anisotropy with respect the one found in the experiment. On 
the other hand, considering nematic pairing, $g_X/g_Y=21$, with $g_X$=5.88 eV$ >  
g_Y=0.28$ eV we reproduce both the absolute values and the angular modulation 
of the observed gap at $\G$ and $X$. For sake of completeness in 
Fig.4 we show the results obtained including the $xy$-pairing channel. We use 
$g_{xy}= -0.45$ eV in order to reproduce the experimental data from ref. 
\onlinecite{Sprau_Science2017}. The inclusion of such contribution does not affect the 
discussion above. Notice that our results are rather robust with respect to 
small variations of the band parameters and nematic splitting. For example, a 
larger nematic splitting $\Delta \Sigma_e=20$ at $M$ as initially estimated in 
ref.\ \onlinecite{Fanfarillo_PRB2016}, would not alter our results, see Fig.\ 
\ref{figlarge}. The only difference is that in this case one has a larger gain 
from the orbital ordering at $M$, so that the experimental data are reproduced 
with a slightly smaller nematic pairing $g_X/g_Y=19$.

\begin{table}[tbh]
\begin{center}
\begin{tabular}{ccccccccccccccccccc}
\hline
& \quad$\Delta^{yz}_h$ \quad &  \quad$\Delta^{xz}_h$  \quad&  \quad$\Delta^{yz}_e$ \quad 
&  \quad$\Delta^{yz}_e$   \quad&  \quad$\Delta^{xy}_X$ \quad & \quad $\Delta^{xy}_Y$ \\
\hline\hline
& \quad$15.71$ \quad &  \quad$0.19$  \quad&  \quad$-1.46$ \quad &  
\quad$-0.71$   \quad&  \quad$-0.20$ \quad & \quad $-0.23$ \\
\hline
\end{tabular}
\caption{$T=0$ orbital SC order parameters (in meV) of the band gap shown in Fig 4 
of the main text.} 
\label{orb_gap}
\end{center}
\end{table} 

It is interesting to compare the value of the anisotropy 
obtained here with the anisotropy of the OSSF extracted from the analysis of the 
shrinking effect in ref.\ \onlinecite{Fanfarillo_PRB2016}. In ref.\ \onlinecite{Fanfarillo_PRB2016} 
the spectral function of the spin modes along the two directions has been 
modelled as: 
\be
\lb{bom}
B_{X/Y}(\omega)=\frac{1}{\pi}\frac{ \omega \omega_0}{\omega_{X/Y}^2+\omega^2}.
\ee
While at RPA level the spin modes are always degenerate, taking into account 
spin-spin interactions beyond Gaussian level 
\cite{Fernandes_PRB2012,Fanfarillo_PRB2018} one can show that below $T_S$ spin 
fluctuations break the $Z_2$ Ising degeneracy and they become stronger at a 
given $\bQ$ vector. This is encoded in Eq.\ \pref{bom} above with two 
anisotropic masses $\omega_{X/Y}$ below $T_S$. The analysis of the orbital 
ordering induced by OSSF discussed in \cite{Fanfarillo_PRB2016} and outlined above 
shows that below $T_S$ one should then have stronger spin fluctuations at 
$\bQ_X$, which implies $\omega_X<\omega_Y$ and $V_X>V_Y$, where $V_{X/Y}$ is 
the coupling of the fermions to the spin modes. The strength of the pairing 
interaction is given by the product of the real part $\chi'_{X/Y}(\omega=0)$ of the 
spin-fluctuation propagator at $\omega=0$ times the spin-mode coupling $V_{X/Y}$. 
Using then the Kramers-Kronig relation of $\chi'$ to the spectral function 
\pref{bom} we get: 
\bea
g_{X/Y}&\propto& V_{X/Y} \chi'_{X/Y}(\omega=0)=V_{X/Y} \int d\omega \frac{B_{X/Y}(\omega)}{\omega}=\nn\\
&=&V_{X/Y}\frac{\omega_0}{\omega_{X/Y}}
\eea
The analysis of ARPES measurements performed in ref.\ 
\onlinecite{Fanfarillo_PRB2016} gives at $T>T_c$ $\omega_Y/\omega_X=1.6$ and 
$V_X/V_Y=8$. Here we used, in the notation of ref.\ 
\onlinecite{Fanfarillo_PRB2016}, the values of the coupling $V^{eh}_{X/Y}$. With 
these numbers we obtain $g_X/g_Y=13.4$, a ratio of the same order of magnitude 
of what we extracted from the gap anisotropy $g_X/g_Y=21$. Notice that the present estimate 
does not take into account the feedback of the SC order on the spin 
modes and could explain the difference between the two results. Such a full 
self-consistent treatment is beyond the scope of the present manuscript and will 
be addressed in a future work. 

Let us finally address the issue of the $k_z$ gap dependence. As discussed in
the main text, a crucial difference when moving from $k_z=0$ to $k_z=\pi$ is
that the orbital character of the hole pocket changes considerably. Since all
the FS pockets expand \cite{Xu_PRL2016, Hashimoto_NatCom2018, Kushirenko_PRB2018},
the effect of the nematic order is less dramatic on the hole pocket, with the
consequence that it retains full $yz$ character at $\theta=0$
\cite{Hashimoto_NatCom2018}. This has already a profound impact on the hole-gap
anisotropy, as recently pointed out in ref.\ \onlinecite{Kang_PRL2018}. To
highlight the effect of the change of orbital weights on the hole pocket at $Z$
we analyze the BCS solution using a set of realistic band parameters for the
Z-pocket as listed in Table \ref{bpz}.
\begin{table}[tbh]
\begin{center}
\begin{tabular}{ccccccccccccccccccc}
\hline 
&              &    $Z$   \\
\hline \hline
&$\epsilon_Z$ &               &26\\
\hline
& $a_Z$       &               &473\\
\hline
& $b_Z$       &               &264\\
\hline
\hline
\end{tabular}
\caption{Low-energy model parameters for the hole-pocket at $k_z=\pi$ in the nematic phase.}
\label{bpz}
\end{center}
\end{table} 

\begin{figure}[t]
\includegraphics[angle=0,width=0.98\linewidth]{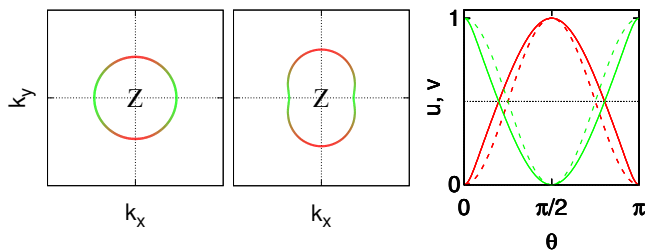} \caption{
$k_z = \pi$ cut of the FS FeSe in the (a) paramagnetic and (b) nematic phase. 
The colors represent the main orbital character. (c) Orbital component amplitude 
of the FS as a function of the azimuthal angle theta in the tetragonal (dashed lines) 
and nematic phase (solid lines)} 
\label{figZ}
\end{figure}

In the absence of a detailed comparison with the band structure above $T_S$ as
done in ref.\ \onlinecite{Fanfarillo_PRB2016}, we already include in $\epsilon_Z$ the
effect of the isotropic shrinking, $(\Sigma^\G_{yz}+\Sigma^\G_{xz})/2$, and
consider separately a further nematic splitting $\Delta\Sigma_h=10$ meV. In
Fig.\ \ref{figZ} we show the FS shape and composition of the Z pocket in both
the paramegnetic and nematic phase. Notice that even below $T_s$, the large
elliptical Z pocket, $k_F^x=0.10$ \AA$^{-1}$ and $k_F^y=0.15$ \AA$^{-1}$,
retains a marked $yz$ character at $\theta=0$ in agreement with the ARPES
measurements \cite{Xu_PRL2016, Hashimoto_NatCom2018, Kushirenko_PRB2018}. Moreover
given this band structure, the inclusion of a finite spin-orbit coupling does
not lead to any significant change, and the $u_Z$, $v_Z$, factors defined as
Eq.s \pref{uvG}, \pref{uvX} scale approximately as $\cos\theta$ and $\sin\theta$
even below $T_s$.

Solving the BCS equation in this case one can easily check that, contrary to
what found at $k_z=0$,  $\langle u_Z^4\rangle \sim \langle v_Z^4\rangle $, so
that the increase of the $\langle u^4_X\rangle$ factor over the $\langle
u^4_Y\rangle$ in the nematic phase is enough to have
$\Delta_h^{yz}>\Delta_h^{xz}$. In this case already without nematic pairing one
has a larger gap value at $\theta=0$, as observed indeed in ref.\ \onlinecite{Kang_PRL2018} (see also Fig 5 in the main text).
As a consequence, if one retains instead the pairing anisotropy  $g_X/g_Y$ 
extracted from the fitting of the $k_z=0$ gaps the anisotropic effect due to the nematic pairing 
is amplified now by the orbital factors that cooperates to give the same gap modulation. As a result
we would get larger gap values and larger anisotropy when moving from the $\G$ to the 
$Z$ pocket. 
Recent ARPES experiments investigated the SC gaps at $k_z=\pi$ 
\cite{Xu_PRL2016, Kushirenko_PRB2018, Hashimoto_NatCom2018, Rhodes_arxiv2018}. While all 
the experiments confirm the in-plane anisotropy of the gap of the hole-pocket, 
with a larger gap value at $\theta=0$, the various reports are somehow in 
disagreement on the $k_z$ dependence of the gap-magnitude. In fact in ref.s\ 
\onlinecite{Xu_PRL2016, Kushirenko_PRB2018} the absolute value of the gap and its 
anisotropy are found to be larger at $k_z=\pi$, while in ref.s \onlinecite{Hashimoto_NatCom2018, Rhodes_arxiv2018} 
the authors claimed a decreasing of the 
gap magnitude when moving from the $\Gamma$ to the $Z$ pocket. The present 
situation calls for a more detailed experimental analysis and specific 
theoretical studies involving a 3D modeling of the band structure. 
As a matter of fact, details of the 3D band model could change 
the estimate of the magnitude of the gaps at different $k_z$, influencing the 
orbital ordering effects and the balance between such mechanism and the nematic 
pairing one. For example, in ref.\ \onlinecite{Kang_PRL2018} the authors obtain a 
larger anisotropy $\Delta_{max}/\Delta_{min}$ at the Z pocket than in our case. 
This is possibly due to a much larger suppression of $xz$ character at the $Y$ 
pocket in their model, leading to a larger $\langle u_X^4\rangle \gg \langle 
u_Y^4\rangle $ anisotropy at the electron pockets. Unfortunately, the 
controversy on the observation of the $Y$ pocket in 
ARPES \cite{Kushirenko_PRB2018, Rhodes_arxiv2018} does not allow us to disentangle this 
issue experimentally. Finally, notice that the authors of ref.\ 
\onlinecite{Kang_PRL2018} do not solve the self-consistent equations at $T=0$, as 
we do, but the linearized ones near $T_c$. Since in a multiband system the 
ratios of the gaps in the various bands depend on temperature, one cannot 
trivially compare those results with the one discussed in the present 
manuscript. Nonetheless, the main qualitative findings, and in particular the 
wrong gap anisotropy found at $\G$ without nematic pairing, hold in both works, 
apart from possible quantitative differences.

\end{document}